\documentclass[12pt]{article}
\usepackage{epsfig}

\def\dfrac#1#2{{\displaystyle {#1 \over #2}}}

\newcommand{\beq}{\begin{equation}}
\newcommand{\eeq}{\end{equation}}
\newcommand{\bea}{\begin{eqnarray}}
\newcommand{\eea}{\end{eqnarray}}
\newcommand{\mev}{\ {\rm MeV}}
\newcommand{\gev}{\ {\rm GeV}}
\newcommand{\ri}{{\rm RI/MOM}}
\newcommand{\msbar}{{\overline{\rm MS}}}
\newcommand{\ms}{m_s}
\newcommand{\ml}{m_\ell}
\begin{document}
\thispagestyle{empty} 
\begin{flushright}
\begin{tabular}{l}
{\tt RM3-TH/02-11}\\
{\tt Roma-1338/02}
\end{tabular}
\end{flushright}
\begin{center}

\vskip 1.2cm
{\centering \LARGE \bf Continuum Determination of Light Quark} \\
\vskip 0.50cm
{\centering \LARGE \bf  Masses from Quenched Lattice QCD}\\

\vskip 1.0cm
{\par\centering \large  
\sc D.~Be\'cirevi\'c$\,^a$, V.~Lubicz$\,^b$ and C.~Tarantino$\,^b$}

\vskip 0.75cm
{\tt {\Large The $\mathrm{SPQ_{CD}R}$ Collaboration}}

\vskip 0.75cm
{\sl 
$^a$ Dip. di Fisica, Universit\`a di Roma ``La Sapienza",\\
Piazzale Aldo Moro 2, I-00185 Rome, Italy. \\                                   
\vspace{.25cm}
$^b$ Dip. di Fisica, Univ. di Roma Tre and INFN,
Sezione di Roma III, \\
Via della Vasca Navale 84, I-00146 Rome, Italy.}\\
%\vskip1.cm
 
\end{center}

\vskip 1.25cm
\begin{abstract}
We compute the strange and the average up/down quark masses in the quenched 
approximation of lattice QCD, by using the ${\cal O}(a)$-improved Wilson action 
and operators and implementing the non-perturbative renormalization. Our 
computation is performed at four values of the lattice spacing, from which we 
could extrapolate to the continuum limit. Our final result for the strange 
quark mass is $\ms^\msbar (2 \ \gev) = (106\pm 2 \pm 8)$~MeV. For the average 
up/down quark mass we obtain $\ml^\msbar (2 \ \gev) =(4.4\pm 0.1 \pm 0.4)$~MeV 
and the ratio $\ms/\ml = (24.3 \pm 0.2 \pm 0.6)$.
\end{abstract}

%\vskip 3.3 cm 
\newpage 
\setcounter{page}{1}
\setcounter{footnote}{0}
\setcounter{equation}{0}
%%%%%%%%%%%%%%%%%%%%%%%%%%%%%%%%%%%%%%%%
%%%%%%%%%%%%%%%%%%%%%%%%%%%%%%%%%%%%%%%%
%%%%%%%%%%%%%%%%%%%%%%%%%%%%%%%%%%%%%%%%
\noindent

\renewcommand{\thefootnote}{\arabic{footnote}}
\vspace*{-1.5cm}

%\newpage
\setcounter{footnote}{0}
%%%%%%%%%%%  Section 1
\section{Introduction}
\setcounter{equation}{0}
In recent years, the determination of quark masses has become one of the main 
research topics of lattice QCD simulations. An accurate determination of these 
masses is in fact of great importance for both phenomenological and theoretical
studies. The masses of the charm and bottom quarks, for instance, enter the 
theoretical predictions of beauty hadron decay rates which, in turn, are 
relevant for the phenomenological analysis of the CKM unitarity triangle and
thus of CP violation in the Standard Model. On the more theoretical side, a 
precise knowledge of quark masses may give insight on the physics of flavour, 
by revealing relations between masses and mixing angles or specific textures of 
the quark mass matrices, eventually due to still uncovered flavour symmetries.

The values of quark masses cannot be directly measured in the experiments, 
since quarks are confined inside the hadrons. On the other hand, being 
fundamental parameters of the theory, quark masses cannot even be computed on 
the basis of purely theoretical considerations. Their evaluation is based on 
the comparison of the theoretical estimate of a physical quantity, which 
depends on quark masses, and its experimental value. This is typically realized 
on the lattice by using, as experimental input, the values of pseudoscalar 
or vector meson masses.

In this paper we present the results of an extensive lattice calculation of 
the strange and the average up/down quark masses in the quenched approximation. 
Particular attention has been dedicated to the reduction and
control of the systematic uncertainties, particularly in the case of the 
strange quark mass. Leading ${\cal O}(a)$ discretization effects (where $a$ is 
the lattice spacing) have been removed by using the non-perturbatively 
${\cal O}(a)$-improved Wilson action and operators~\cite{improvement}. The 
systematic uncertainty related to the evaluation of the quark mass 
renormalization constant in perturbation theory has been significantly reduced, 
by implementing the non-perturbative renormalization technique of 
ref.~\cite{ri-mom} in the $\ri$ scheme. Conversion of the quark masses from the 
$\ri$ scheme to the most popular $\msbar$ scheme has been performed by using 
continuum perturbation theory at the $\mathrm{N^3LO}$~\cite{chetyr}. Finite 
volume effects have also been studied. In order to estimate residual systematic
uncertainties, we have compared the results obtained by using two alternative 
definitions of the lattice bare quark mass, related to the vector and 
axial-vector Ward identities respectively. Finally, with respect to previous 
determinations of light quark masses by our collaboration~\cite{bec1,bec2}, 
in this study we have performed a calculation at four different values of the 
lattice scale, corresponding to an inverse lattice spacing in the range between 
approximately 2 and 4 GeV. In this way, we have been able to extrapolate our 
results to the continuum limit. The main relevant source of uncertainty, which 
is left in our calculation of the strange quark mass, is therefore the quenched 
approximation.

An additional uncertainty is present in the determination of the average 
up/down quark mass. Typical values of the lightest quark masses, used in the 
present and most of current lattice calculations, are approximately of the 
order of $\ms/2$, where $\ms$ is the strange quark mass. Therefore, a large 
chiral extrapolation is required to reach the physical values of the up and 
down quark masses. Chiral perturbation theory may be used as a guidance in this 
extrapolation, but the inclusion of higher order terms in the chiral expansion,
necessary to increase the accuracy of this determination, requires simulations 
with many more (and preferably lighter) quark masses . In the region of masses 
considered in this paper, the pseudoscalar meson mass squared shows a good 
linear dependence on quark masses, and a linear or quadratic fit has been 
considered in performing the chiral extrapolation. The systematic uncertainty 
introduced by this extrapolation is difficult to be reliably quantified. It 
will be assessed only when simulations on larger lattice volumes and smaller 
values of quark masses become feasible. We stress, however, that this 
uncertainty does not affect the determination of the strange quark mass.

We conclude this section by summarizing the main results of this paper. For the 
strange quark mass and the average value of the up/down quark masses, 
$\ml=(m_u+m_d)/2$, quoted in the $\msbar$ scheme at the renormalization scale 
$\mu=2 \gev$, we obtain
\beq
\ms^{\msbar}(2 \gev)= (106\pm 2 \pm 8)\ \mev
\eeq
and
\beq
\ml^{\msbar}(2 \gev)= (4.4\pm 0.1 \pm 0.4) \ \mev \, ,
\eeq
in good agreement with the current lattice world averages~\cite{lubicz,Kaneko}.
For the ratio of the strange to the average light quark mass we find
\beq
\frac{\ms}{\ml}=24.3 \pm 0.2 \pm 0.6 \,.
\eeq
This result is in good agreement with the prediction $\ms/\ml=24.4\pm 1.5$ 
based on NLO chiral perturbation theory~\cite{leutwyler}. 

\section{Details of the lattice calculation}
In determining the values of quark masses we used the standard procedures based
on the vector and axial-vector Ward identities~\cite{boch}. 

\vspace{0.5cm}
\noindent
{\em Vector Ward Identity} (VWI): in the renormalized 
continuum theory the VWI reads
\beq\label{e1}
\langle \partial_\mu V_\mu(x) {\cal
O}^\dagger(0)\rangle = (m_1(\mu) -m_2(\mu)) \langle S(\mu;x) {\cal
O}^\dagger(0)\rangle\;,
\eeq
where $m_{1,2}$ are the quark masses, $V_\mu = \bar q_1 \gamma_\mu q_2$ is the
vector current, and $S = \bar q_1 q_2$ is the scalar density. The renormalized 
quark mass, $m_q(\mu)= Z_m(\mu\,a)\, m_q(a)$, is obtained from the bare mass 
$m_q(a)$ which, on the lattice with Wilson fermions, is equal to $m_q(a) = (1/2
\,a)(1/\kappa_q -  1/\kappa_{cr})$. $\kappa_q$ is the Wilson hopping parameter 
and $\kappa_{cr}$ is its critical value, corresponding to the chiral limit. The 
VWI relates the quark mass renormalization constant to the one of the scalar 
density, {\it i.e.} $Z_m= Z_S^{-1}$. The computation of the quark mass, using 
the VWI, can be summarized by the following formula:
\beq
\label{VWI}
m_q^{\rm (VWI)}(\mu) = Z_m(\mu\,a) m_q^{\rm (VWI)}(a) = 
Z_S^{-1}(\mu\,a) \biggl( 1 + b_m\,a m_q \biggr) m_q(a) \ .
\eeq
Notice that the parameter $b_m=-b_S/2$ provides the elimination of ${\cal O}
(am_q)$ effects~\cite{improvement}.

\vspace{0.5cm}
\noindent
{\em Axial-Vector Ward Identity} (AWI): the renormalized continuum AWI reads 
\beq
\label{awi}
\langle \partial_\mu A_\mu(x) {\cal
O}^\dagger(0)\rangle = 2 m(\mu) \langle P(\mu;x) {\cal
O}^\dagger(0)\rangle
\eeq
where $A_\mu = \bar q \gamma_\mu \gamma_5 q$ is the axial current, $P = \bar q 
\gamma_5 q$ is the pseudoscalar density, ${\cal O}$ is a generic operator and 
we have considered quark fields with degenerate masses. The quark mass 
renormalization constant, in this case, is related by the AWI to the 
renormalization constant of the axial and of the pseudoscalar operators, 
{\it i.e.} $\bar Z_m=Z_A/Z_P$. By choosing ${\cal O} = P$ we have:
\beq
\label{AWI}
m_q^{\rm (AWI)}(\mu\,a) = \bar Z_m(\mu\,a) m_q^{\rm (AWI)}(a) = 
{Z_A \over Z_P(\mu\,a)}{ \biggl( 1 + (b_A - b_P )\, a m_q \biggr)} 
{ \langle \displaystyle{\sum_{\vec x}} \partial_0 A^I_0
(x) P^\dagger(0)\rangle \over 2\  \langle 
\displaystyle{\sum_{\vec x}} P (x) P^\dagger(0)\rangle }  \ ,
\eeq
where the bare axial current is improved at ${\cal O}(a)$ as $A^I_\mu(x) = 
A_\mu(x) + a\,c_A \partial_\mu P(x)$. The coefficient ($b_A - b_P$) cancels the 
terms of ${\cal O}(am_q)$. For the time derivative, we consider the symmetric 
(${\cal O}(a)$-improved) form, {\it i.e.} $\partial_0 f(t) =(f(t+a)-f(t-a))/2a$.
The improvement coefficients $b_m$, $b_A-b_P$ and $c_A$, in eqs.~(\ref{VWI}) 
and (\ref{AWI}), are only functions of the bare lattice coupling $g_0^2$. 

\vspace{0.5cm}
Complete information about the lattice calculation performed in this study is 
provided in table~\ref{tab1}. We generated ${\cal O}(1000)$ gauge field
configurations in the quenched approximation at four values of the gauge 
coupling constant, corresponding to an inverse lattice spacing in the range 
between approximately 2 and 4 GeV. For each value of the lattice spacing, quark 
propagators have been computed at four light values of the bare quark mass, by
using the non-perturbatively ${\cal O}(a)$-improved Wilson 
action~\cite{improvement,alphaCSW}.
%%%%%%%%%%%%%%%%%%%%%%%%%%%%%%%%%%%%%%%%%%%%%%%%%%%%%%%%%%%%%%%%%%%%%%%%%%%%%%%
\begin{table}
\centering 
\begin{tabular}{|c|cccc|}  \hline \hline
{\phantom{\huge{l}}}\raisebox{-.2cm}{\phantom{\Huge{j}}}
$ \beta = 6/g_0^2$&  6.0 &  6.2 & 6.4   & 6.45    \\ 
{\phantom{\huge{l}}}\raisebox{-.2cm}{\phantom{\Huge{j}}}
$ c_{SW}$~\cite{alphaCSW} &   1.769 &  1.614 & 1.526  & 1.509 \\ 
{\phantom{\huge{l}}}\raisebox{-.2cm}{\phantom{\Huge{j}}}
$ L^3 \times T $&  $16^3 \times 52$ & $24^3 \times 64$  & $32^3 \times 70$& $32^3 \times 70$ \\ 
{\phantom{\huge{l}}}\raisebox{-.2cm}{\phantom{\Huge{j}}}
$ \#\ {\rm conf.}$& 500 &  200 & 150  & 100\\
{\phantom{\huge{l}}}\raisebox{-.2cm}{\phantom{\Huge{j}}}
$a^{-1}$(GeV) & 2.00(10) & 2.75(14) & 3.63(18) & 3.87(19) \\
%{\phantom{\huge{l}}}\raisebox{-.2cm}{\phantom{\Huge{j}}}
%$a^{-1}(m_{K^\ast})$ & 2.10(8) GeV& 2.58(8) GeV & 3.39(6) GeV  & 3.74(16) GeV \\
%{\phantom{\huge{l}}}\raisebox{-.2cm}{\phantom{\Huge{j}}}
%$a/r_0$~\cite{necco} &   0.1863 & 0.1354 & 0.1027 & 0.0962 \\  
\hline 
{\phantom{\huge{l}}}\raisebox{-.2cm}{\phantom{\Huge{j}}}
$\kappa_1$& 0.1335 & 0.1339 & 0.1347 & 0.1349   \\ 
{\phantom{\huge{l}}}\raisebox{-.2cm}{\phantom{\Huge{j}}}
$\kappa_2$& 0.1338 & 0.1344 &  0.1349 &  0.1351  \\ 
{\phantom{\huge{l}}}\raisebox{-.2cm}{\phantom{\Huge{j}}}
$\kappa_3$& 0.1340 & 0.1349 &  0.1351 &  0.1352  \\ 
{\phantom{\huge{l}}}\raisebox{-.2cm}{\phantom{\Huge{j}}}
$\kappa_4$& 0.1342 & 0.1352 &  0.1353 &  0.1353  \\
{\phantom{\huge{l}}}\raisebox{-.2cm}{\phantom{\Huge{j}}}
$\kappa_{cr}$&   0.135175(4)& 0.135785(2) & 0.135734(2) & 0.135680(2) \\  \hline 
{\phantom{\huge{l}}}\raisebox{-.2cm}{\phantom{\Huge{j}}}
$Z_{A}$~\cite{RCS-Roma} &  0.804(2) & 0.809(2) & 0.824(2) & 0.825(4)\\
{\phantom{\huge{l}}}\raisebox{-.2cm}{\phantom{\Huge{j}}}
$Z^{RI}_{S}(\mu = 3$ GeV)~\cite{RCS-Roma} &  0.745(3) & 0.692(3) & 0.668(4) & 0.668(7)\\
{\phantom{\huge{l}}}\raisebox{-.2cm}{\phantom{\Huge{j}}}
$Z^{RI}_{P}(\mu = 3$ GeV)~\cite{RCS-Roma} &  0.598(3) & 0.575(4) & 0.576(4) & 0.579(8)\\
{\phantom{\huge{l}}}\raisebox{-.2cm}{\phantom{\Huge{j}}}
$c_A$~\cite{alphaCSW,LANL}& -0.038
%$\begin{array}{c}-0.038\\-0.083\end{array}$ 
& -0.038   & -0.025 & -0.023 \\ 
{\phantom{\huge{l}}}\raisebox{-.2cm}{\phantom{\Huge{j}}}
$b_m$~\cite{alphabM} & -0.709& -0.691 & -0.676 & -0.673  \\
{\phantom{\huge{l}}}\raisebox{-.2cm}{\phantom{\Huge{j}}}
$(b_A-b_P)$~\cite{alphabM} & 0.171&  0.039 & 0.013 & 0.010 \\  \hline
{\phantom{\huge{l}}}\raisebox{-.2cm}{\phantom{\Huge{j}}}
$am_P$: fit for $t\in $ & $[11\div 25]$ & $[12\div 31]$ & $[17\div 34]$ & $[17\div 34]$ \\
{\phantom{\huge{l}}}\raisebox{-.2cm}{\phantom{\Huge{j}}}
$am_V$: fit for $t\in $ & $[11\div 23]$ & $[12\div 28]$ & $[17\div 28]$ & $[17\div 30]$ \\
{\phantom{\huge{l}}}\raisebox{-.2cm}{\phantom{\Huge{j}}}
$(am_q)^{\rm AWI}$: fit for $t\in $ & $[11\div 24]$ & $[13\div 29]$ & $[17\div 32]$ & $[17\div 31]$\\
\hline \hline
\end{tabular}
{\caption{\sl \small \label{tab1} Summary of the lattice details and parameters 
used in this work. We also give the values of the inverse lattice spacing, of 
the critical hopping parameter and of the renormalization constants and 
improvement coefficients (with corresponding references). In addition we supply 
the reader with the fit intervals that have been used for all the correlation 
functions considered in this work. Note that our time counting is $0,\dots, 
(T-1)$. }}
\end{table}
%%%%%%%%%%%%%%%%%%%%%%%%%%%%%%%%%%%%%%%%%%%%%%%%%%%%%%%%%%%%%%%%%%%%%%%%%%%%%%%

In view of the final extrapolation of the lattice results to the continuum 
limit, an important requirement in this calculation is a precise determination 
of the lattice spacing which corresponds to the different values of the 
coupling used in this study. While the absolute value of the physical scale is 
affected by a rather large uncertainty in the quenched approximation (of the 
order of 10\%), the ratio between two scales can be determined with better 
accuracy. To this purpose, we use the precise determination based on the study 
of the static quark anti-quark potential~\cite{necco}, which in the range $5.7 
\le \beta \le 6.92$ can be expressed in the form
\beq
\label{necco}
\ln\left(\dfrac{a^{-1}(\beta)}{a^{-1}(\beta=6)}\right) = 1.7331\,(\beta-6) -
0.7849\,(\beta-6)^2 + 0.4428\,(\beta-6)^3 \, .
\eeq
By using this formula, and varying the inverse lattice spacing at the reference 
point $\beta=6.0$ in the conservative range between 1.9 and 2.1 GeV, we obtain 
the estimates of the scale given in table~\ref{tab1}.

In the same table, we also give the results for the relevant renormalization 
constants, $Z_A$, $Z_S$ and $Z_P$, in the chiral limit, which have been 
determined by using the non-perturbative renormalization method of 
ref.~\cite{ri-mom}, in the $\ri$ scheme. The scale dependent constants, $Z_S$ 
and $Z_P$, have been computed at the scale $\mu=3 \gev$, which lies in the 
allowed range $\Lambda_{QCD} < \mu < \pi/a$ for all the values of the coupling 
considered in this study. For this reason, our non-perturbative results for the
quark masses in the $\ri$ scheme will be given at the reference scale $\mu=3
\gev$. Details of the non-perturbative calculation of the renormalization 
constants have been presented at the ``Lattice 2002" conference and will be 
discussed in a separate publication~\cite{RCS-Roma}.

Concerning the values of the improvement coefficients, $c_A$, $b_m$ and 
($b_A - b_P$), we use the non-perturbative determinations of 
refs.~\cite{alphaCSW,LANL,alphabM}, whose results are collected in 
table~\ref{tab1}. Notice that at $\beta=6.0$ we opted for the value of $c_A$ 
obtained in~\cite{LANL} (and recently confirmed in~\cite{glasgow}), whose 
absolute value is significantly smaller than the one obtained in 
ref.~\cite{alphaCSW}. Had we used the value of $c_A$ obtained in 
ref.~\cite{alphaCSW}, we would have found values of the AWI quark masses, at 
$\beta=6.0$, smaller by approximately 7\%.

In table~\ref{tab2}, we show the results for the pseudoscalar and vector meson
masses, in lattice units, obtained by fitting the corresponding correlation 
functions at zero spatial momentum in the time intervals indicated in 
table~\ref{tab1}. For each value of the hopping parameter, we also present in 
table~\ref{tab2} the corresponding values of the VWI and AWI quark masses, 
defined in eqs.~(\ref{VWI}) and (\ref{AWI}), renormalized in the $\ri$ scheme 
at the scale $\mu = 3\,\gev$. 
%%%%%%%%%%%%%%%%%%%%%%%%%%%%%%%%%%%%%%%%%%%%%%%%%%%%%%%%%%%%%%%%%%%%%%%%%%%%%%%
\begin{table}
\begin{center} 
\hspace*{-5mm}\begin{tabular}{|c|c|c|c|c|c|} 
\hline
{\phantom{\Huge{l}}}\raisebox{-.2cm}{\phantom{\Huge{j}}}
{$\beta\qquad$}& { $\qquad \kappa \qquad $ }  & 
$\qquad a m_P\qquad $  & $\qquad a m_V\qquad$ & 
$a m_q^{\rm (VWI)}$ & $a m_q^{\rm (AWI)}$  \\   \hline \hline 
{\phantom{\Huge{l}}}\raisebox{-.2cm}{\phantom{\Huge{j}}}
$\mathsf{ 6.0}\qquad$   & 0.1335 &   0.391(1) & 0.524(4)& 0.0602(3)& 0.0672(4)  \\ 
{\phantom{\Huge{l}}}\raisebox{-.2cm}{\phantom{\Huge{j}}}
                     & 0.1338 &   0.356(1) & 0.498(6)& 0.0496(2)& 0.0554(3)  \\ 
{\phantom{\Huge{l}}}\raisebox{-.2cm}{\phantom{\Huge{j}}}
                     & 0.1340 &   0.331(1) & 0.480(7)& 0.0425(2)& 0.0475(3)  \\ 
{\phantom{\Huge{l}}}\raisebox{-.2cm}{\phantom{\Huge{j}}}
                     & 0.1342 &   0.304(1) & 0.462(9)& 0.0353(2)& 0.0396(2)  \\ \hline
{\phantom{\Huge{l}}}\raisebox{-.2cm}{\phantom{\Huge{j}}}
$\mathsf{ 6.2}\qquad$ & 0.1339 &   0.357(1) & 0.443(3)& 0.0722(3)& 0.0781(4)  \\ 
{\phantom{\Huge{l}}}\raisebox{-.2cm}{\phantom{\Huge{j}}}
                 & 0.1344 &   0.303(1) & 0.405(4)& 0.0534(2)& 0.0575(3)  \\ 
{\phantom{\Huge{l}}}\raisebox{-.2cm}{\phantom{\Huge{j}}}
                 & 0.1349 &   0.243(1) & 0.370(7)& 0.0343(1)& 0.0370(2)  \\ 
{\phantom{\Huge{l}}}\raisebox{-.2cm}{\phantom{\Huge{j}}}
                 & 0.1352 &   0.200(1) & 0.351(11)& 0.0228(1)& 0.0247(1)  \\ \hline
{\phantom{\Huge{l}}}\raisebox{-.2cm}{\phantom{\Huge{j}}}
$\mathsf{ 6.4}\qquad$ & 0.1347 
                      &   0.228(1) & 0.306(2)& 0.0416(3)& 0.0440(3)  \\ 
{\phantom{\Huge{l}}}\raisebox{-.2cm}{\phantom{\Huge{j}}}
             & 0.1349 &   0.204(2) & 0.291(2)& 0.0336(2)& 0.0355(2)  \\ 
{\phantom{\Huge{l}}}\raisebox{-.2cm}{\phantom{\Huge{j}}}
             & 0.1351 &   0.178(2) & 0.277(3)& 0.0256(2)& 0.0271(2)  \\ 
{\phantom{\Huge{l}}}\raisebox{-.2cm}{\phantom{\Huge{j}}}
             & 0.1353 &   0.148(2) & 0.266(4)& 0.0176(1)& 0.0186(1)  \\ \hline
{\phantom{\Huge{l}}}\raisebox{-.2cm}{\phantom{\Huge{j}}}
$\mathsf{ 6.45}\qquad$ & 0.1349 
                      &   0.195(2) & 0.272(4)& 0.0314(3)& 0.0332(4)  \\ 
{\phantom{\Huge{l}}}\raisebox{-.2cm}{\phantom{\Huge{j}}}
             & 0.1351 &   0.167(2) & 0.255(5)& 0.0234(2)& 0.0247(3)  \\ 
{\phantom{\Huge{l}}}\raisebox{-.2cm}{\phantom{\Huge{j}}}
             & 0.1352 &   0.152(2) & 0.247(6)& 0.0194(2)& 0.0205(2)  \\ 
{\phantom{\Huge{l}}}\raisebox{-.2cm}{\phantom{\Huge{j}}}
             & 0.1353 &   0.134(3) & 0.240(8)& 0.0154(1)& 0.0162(2)  \\ \hline
\end{tabular} 
%\vspace*{.8cm}
\caption{\label{tab2}
\small{\sl Pseudoscalar and vector meson masses together with the corresponding
VWI and AWI quark masses, in lattice units. Quark masses are renormalized in 
the $\ri$ scheme at the scale $\mu=3$ GeV. Information about the fit intervals,
the values of the renormalization constants and of the improvement coefficients
can be found in table~\ref{tab1}.}}
\end{center}
\end{table}
%%%%%%%%%%%%%%%%%%%%%%%%%%%%%%%%%%%%%%%%%%%%%%%%%%%%%%%%%%%%%%%%%%%%%%%%%%%%%%%

\begin{figure}
%\vspace*{-1.cm}
%\begin{center}
\hspace*{-1.cm}
\begin{tabular}{cc}
\epsfxsize8.15cm\epsffile{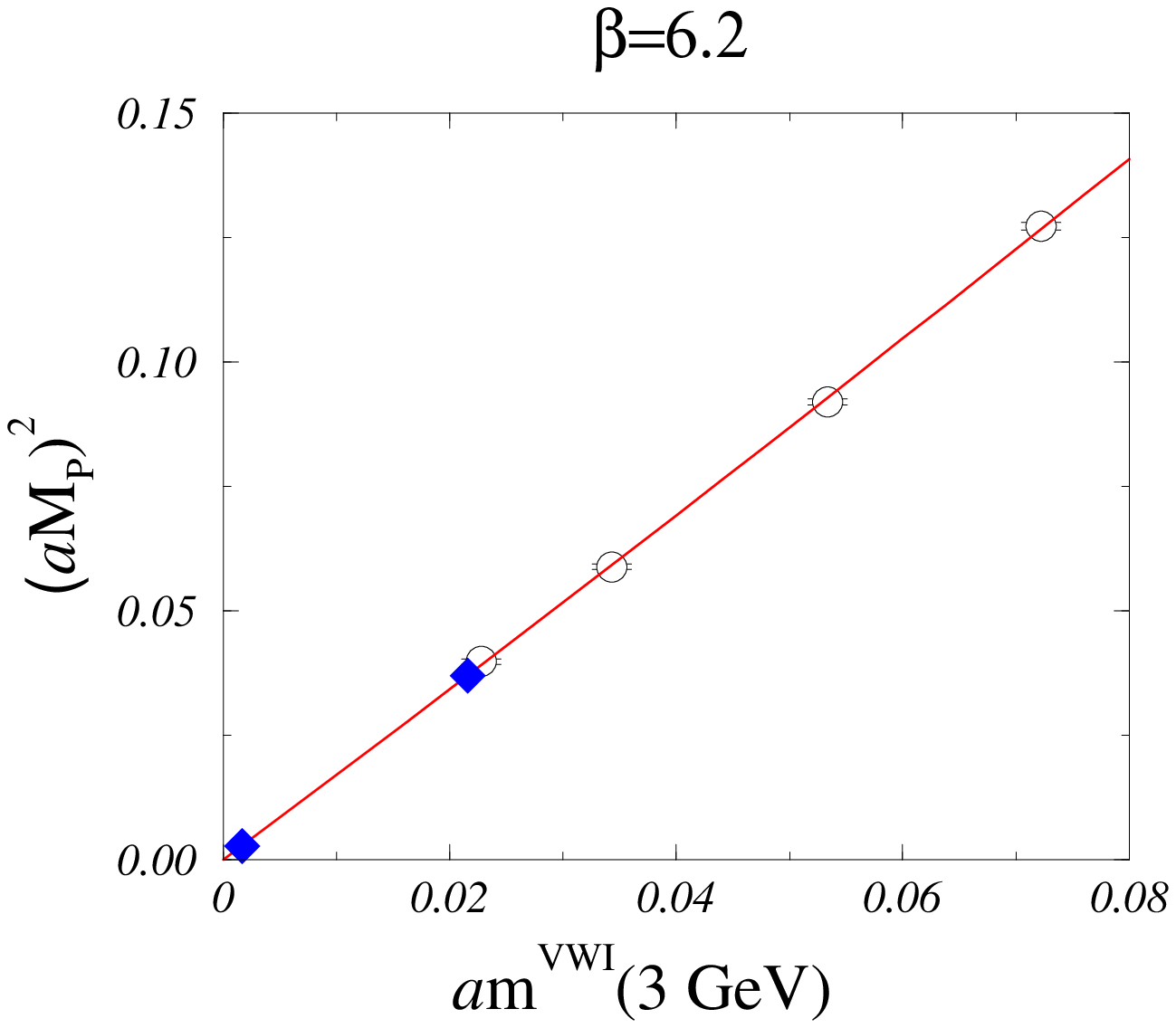} &
\epsfxsize7.93cm\epsffile{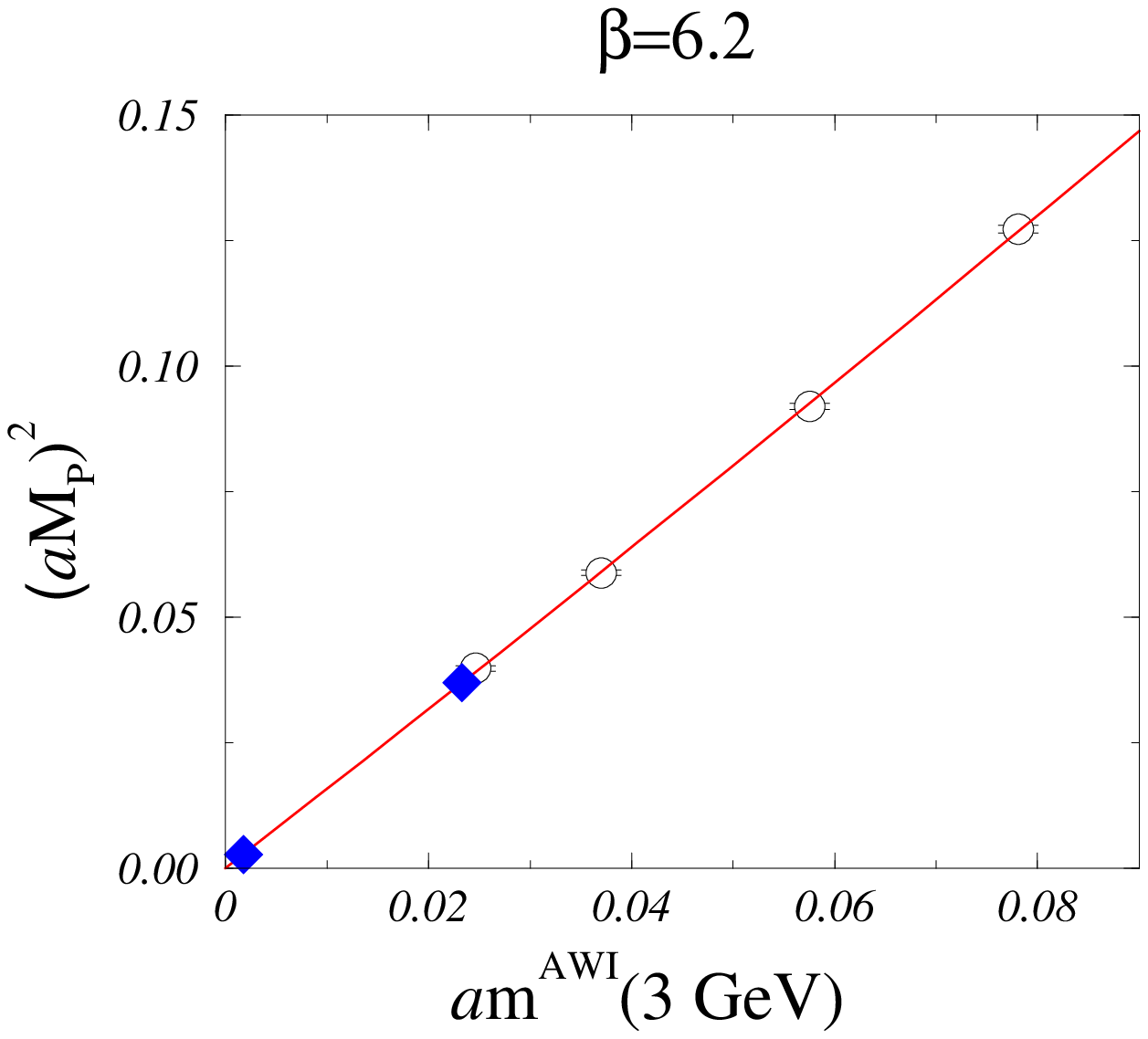} \\
\end{tabular}
\caption{\label{fig:extrapo}{\sl Quadratic fits of the pseudoscalar meson 
mass squared as a function of the renormalized VWI (left) and AWI (right) 
quark masses, at $\beta=6.2$. Empty circles represent the lattice data, full 
diamonds show the physical values of the pion and kaon masses.}}
%\end{center}
\end{figure}

In order to get the physical values of quark masses, we follow the usual 
procedure~\cite{bec1} and fit the pseudoscalar meson masses to the following 
form
\beq 
\label{mpfit}
(a m_{P})^2 = Q_1 \left(a m_1^{\rm (VWI)} + a m_2^{\rm (VWI)}\right) + 
Q_2 \left(a m_1^{\rm (VWI)} + a m_2^{\rm (VWI)}\right)^2\;,
\eeq
and similarly for the AWI quark masses. The subscripts $1$ and $2$ denote the 
flavour of the two valence quarks in the meson. Since in this study we only 
considered mesons consisting of two degenerate quarks, we always have $m_1=m_2$ 
in the fits. For the same reason, we did not include a term proportional to 
$(m_1 - m_2)^2$ in eq.~(\ref{mpfit}). The physical values of the average 
up/down and of the strange quark masses are then obtained by substituting the 
experimental pion and kaon masses on the l.h.s. of eq.~(\ref{mpfit}) and the 
values of the lattice spacing listed in table~\ref{tab1}. Notice that we do not
distinguish the up from the down quark mass and, by using eq.~(\ref{mpfit}), we 
can only determine the average of the two, {\it i.e.} $\ml = (m_u + m_d)/2$. 
The fit of the pseudoscalar meson masses to eq.~(\ref{mpfit}), at $\beta=6.2$, 
and the resulting extrapolation to the physical values, is shown in 
figure~\ref{fig:extrapo}, for both the VWI and AWI quark masses. As can be seen 
from the figure, the effect of including a quadratic term in the chiral 
extrapolation of quark masses is rather negligible. This is true for all values 
of the lattice spacing considered in this study. 
%%%%%%%%%%%%%%%%%%%%%%%%%%%%%%%%%%%%%%%%%%%%%%%%%%%%%%%%%%%%%%%%%%
%%%%%%%%%%%%%%%%%%%%%%%%%%%%%%%%%%%%%%%%%%%%%%%%%%%%%%%%%%%%%%%%%%

The results for the strange and the average up/down quark masses, in the $\ri$ 
scheme, at the renormalization scale $\mu=3\gev$, are collected in 
table~\ref{tab3}. Also shown in the table are the values of the ratio $\ms/\ml$
(which is a both scheme and scale independent quantity). Note that this
quantity can be determined with much better accuracy than the mass $\ml$
itself, since statistical and systematic errors largely cancel in the ratio.
We emphasize that the results for quark masses presented in table~\ref{tab3} 
are obtained in a completely non-perturbative way.
%%%%%%%%%%%%%%%%%%%%%%%%%%%%%%%%%%%%%%%%%%%%%%%%%%%%%%%%%%%%%%%%%%
\begin{table}
\begin{center} 
\begin{tabular}{|c||c|c|c|c|}
\hline
{\phantom{\Huge{l}}}\raisebox{-.1cm}{\phantom{\Huge{j}}}
$\beta$ & 6.0 & 6.2 & 6.4 & 6.45 \\ \hline\hline
{\phantom{\Huge{l}}}\raisebox{-.1cm}{\phantom{\Huge{j}}}
$\ms^{\rm VWI}$ [MeV] & 90(1) & 100(1)   &105(3) &104(3) \\
{\phantom{\Huge{l}}}\raisebox{-.1cm}{\phantom{\Huge{j}}}
$\ms^{\rm AWI}$ [MeV] & 101(1) & 108(1)  & 111(3) & 110(4) \\ \hline
{\phantom{\Huge{l}}}\raisebox{-.1cm}{\phantom{\Huge{j}}}
$\ml^{\rm VWI}$ [MeV] & 3.50(6) &4.02(5) &4.21(13) &4.24(23) \\
{\phantom{\Huge{l}}}\raisebox{-.1cm}{\phantom{\Huge{j}}}
$\ml^{\rm AWI}$ [MeV] & 3.93(6) &4.33(6) &4.45(14) &4.47(24) \\ \hline
{\phantom{\Huge{l}}}\raisebox{-.1cm}{\phantom{\Huge{j}}}
$(\ms/\ml)^{\rm VWI}$  & 25.7(1) & 24.90(8) & 25.0(2) & 24.5(6) \\ 
{\phantom{\Huge{l}}}\raisebox{-.1cm}{\phantom{\Huge{j}}}
$(\ms/\ml)^{\rm AWI}$  & 25.6(1) & 24.91(8) & 25.0(2) & 24.5(6) \\ \hline
\end{tabular}
%\vspace*{.8cm}
\caption{\label{tab3}
{\sl \small Values of the strange and the average up/down quark masses in the 
$\ri$ scheme at the scale $\mu=3$ GeV, as obtained from the VWI and the AWI
methods. We also present the values of the scheme and scale independent ratio 
$\ms/\ml$.}}
\end{center}
\end{table}
%%%%%%%%%%%%%%%%%%%%%%%%%%%%%%%%%%%%%%%%%%%%%%%%%%%%%%%%%%%%%%%%%%

\section{Conversion to the $\msbar$ scheme and extrapolation to the continuum 
limit}  
We now convert the $\ri$ quark masses, obtained in the previous section, to the 
popular $\msbar$ scheme, in which the light quark masses are conventionally 
expressed at the scale $\mu = 2$~GeV. That allows to compare our results to the 
results obtained by other lattice groups and to the ones obtained by using QCD 
sum rules. It is only at this stage of the calculation that we are forced to 
introduce perturbation theory. This is because the $\msbar$ scheme, being 
related to dimensional regularization, is defined in perturbation theory only.

The conversion factor from the $\ri$ to the $\msbar$ scheme is conveniently 
calculated by introducing the renormalization group invariant mass, 
$m^{\rm RGI}$, defined by dividing out from the renormalized quark masses the 
perturbative scale dependence
\beq
\label{mrgi}
m^{\rm RGI} = {\ m^{\rm RI}(\mu)\ \over c^{\rm RI}(\mu)} 
= {\ m^\msbar(\mu)\ \over c^\msbar(\mu)} \;.
\eeq
The mass $m^{\rm RGI}$ is, by definition, both renormalization scale and 
renormalization scheme independent. The beta function of QCD and the quark mass 
anomalous dimension, entering the functions $c(\mu)$ in eq.~(\ref{mrgi}), are 
known to 4-loop accuracy, in both the $\ri$~\cite{chetyr} and the $\msbar$ 
schemes~\cite{vermas1}-\cite{vermas2}. From these papers, we extract
%\bea \label{chet}
%&&c^\ri (\mu) = \left({\alpha_s(\mu)\over \pi }\right)^{4/11} \biggl[\  1\  
%+\ 2.02067\  {\alpha_s(\mu)\over \pi }\  +\  14.21925\ 
%\left({\alpha_s(\mu)\over \pi }\right)^2\ +\biggr.
%\nonumber \\ 
%&& \biggl.\hspace*{63mm} 138.30689\ \left({\alpha_s(\mu)\over \pi }\right)^3\biggr]\;,
%&&c^\ri (\mu) = \left({\alpha_s(\mu)\over \pi }\right)^{12/25} \biggl[\  1\  
%+\ 2.34747\  {\alpha_s(\mu)\over \pi }\  +\ 12.0599\ 
%\left({\alpha_s(\mu)\over \pi }\right)^2\ +\biggr.
%\nonumber \\ 
%&& \biggl.\hspace*{63mm} 84.4076\ \left({\alpha_s(\mu)\over \pi }\right)^3
%\biggr]\;,
%\eea
\beq \label{chet}
c^{\rm RI} (\mu) = \left({\alpha_s(\mu)\over \pi }\right)^\frac{12}{25} 
\left[  1  +\ 2.34747\  {\alpha_s(\mu)\over \pi }  +\ 12.0599\ 
\left({\alpha_s(\mu)\over \pi }\right)^2 + \ 84.4076\ \left({\alpha_s(\mu)
\over \pi }\right)^3 \right]\,,
\eeq
and
%\bea \label{MS}
%&&c^\msbar (\mu) = \left({\alpha_s(\mu)\over \pi }\right)^{4/11} \biggl[\  1\  
%+\ 0.68733\  {\alpha_s(\mu)\over \pi }\  +\  1.51211\ 
%\left({\alpha_s(\mu)\over \pi }\right)^2\ +\biggr.
%\nonumber \\ 
%&& \biggl. \hspace*{63mm}  4.05787\ \left({\alpha_s(\mu)\over \pi
%}\right)^3\biggr]\;.
%\eea
\beq \label{MS}
c^\msbar (\mu) = \left({\alpha_s(\mu)\over \pi }\right)^\frac{12}{25} 
\left[  1  +\ 1.01413\  {\alpha_s(\mu)\over \pi }  +\  1.38921\ 
\left({\alpha_s(\mu)\over \pi }\right)^2 + \ 1.09054\ 
\left({\alpha_s(\mu)\over \pi}\right)^3\right]\,,
\eeq
by using $n_F=4$ as the number of active flavours in the range of scales 
between 2 and 3 GeV. Then, by using $\alpha_s(m_Z)=0.118$, we obtain the 
conversion factor
\beq
\label{conv}
R^{(4)} \equiv {m_q^\msbar (2\gev) \over m^{\rm RI}(3\gev)} = 
{c^\msbar (2\gev)\over c^{\rm RI}(3\gev)} \ = 0.918 \; .
\eeq
This result has a $\mathrm{N^3LO}$ accuracy in continuum perturbation theory. 
Therefore, we expect the perturbative error in our determination of the 
$\msbar$ quark masses to be completely negligible.

We use eq.~(\ref{conv}) to convert the results for the $\ri$ quark masses, 
presented in table~\ref{tab3}, to the $\msbar$ scheme. The resulting values 
of $\ml^\msbar(2\gev)$ and $\ms^\msbar(2\gev)$, as obtained from the VWI
and AWI respectively, are shown in table~\ref{tab4}. We do not report in the
table the values of the ratio $\ms/\ml$, presented in table~\ref{tab3}, since 
this quantity is scheme independent.
%%%%%%%%%%%%%%%%%%%%%%%%%%%%%%%%%%%%%%%%%%%%%%%%%%%%%%%%%%%%%%%%%%
\begin{table}
\begin{center} 
\begin{tabular}{|c||c|c|c|c|}
\hline
{\phantom{\Huge{l}}}\raisebox{-.1cm}{\phantom{\Huge{j}}}
$\beta$ & 6.0 & 6.2 & 6.4 & 6.45 \\ \hline \hline
{\phantom{\Huge{l}}}\raisebox{-.1cm}{\phantom{\Huge{j}}}
$\ms^{\rm VWI}$ [MeV] & 82(1) & 92(1)   &96(2) &96(3) \\
{\phantom{\Huge{l}}}\raisebox{-.1cm}{\phantom{\Huge{j}}}
$\ms^{\rm AWI}$ [MeV] & 92(1) & 99(1)   &102(2) & 101(3) \\ \hline
{\phantom{\Huge{l}}}\raisebox{-.1cm}{\phantom{\Huge{j}}}
$\ml^{\rm VWI}$ [MeV] & 3.22(5) &3.69(5) &3.87(12) &3.89(21) \\
{\phantom{\Huge{l}}}\raisebox{-.1cm}{\phantom{\Huge{j}}}
$\ml^{\rm AWI}$ [MeV] & 3.61(6) &3.98(6) &4.09(13) &4.11(22) \\ \hline
\end{tabular}
%\vspace*{.8cm}
\caption{\label{tab4}
{\sl \small Values of the strange and the average up/down quark masses in the 
$\msbar$ scheme at the scale $\mu=2$ GeV.}}
\end{center}
\end{table}
%%%%%%%%%%%%%%%%%%%%%%%%%%%%%%%%%%%%%%%%%%%%%%%%%%%%%%%%%%%%%%%%%%

\begin{figure}[t]
\begin{center}
\begin{tabular}{c}
\hspace*{.25cm}\epsfxsize10.0cm\epsffile{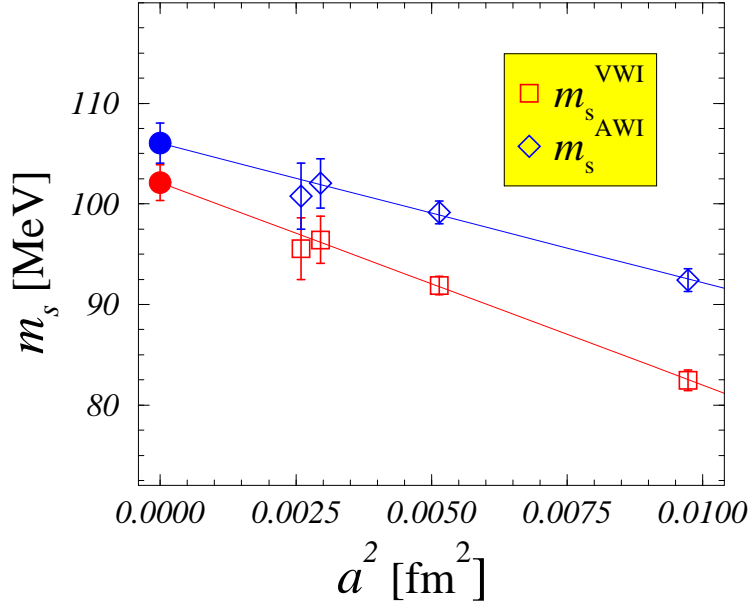}   \\
\end{tabular}
\caption{\label{fig1}{\sl Extrapolation of the $\msbar$ strange quark 
mass, at $\mu=2\gev$, to the continuum limit.}}
\end{center}
\end{figure}

The last step of our calculation is the extrapolation to the continuum limit. 
As discussed in the previous section, we find it convenient for that purpose to 
fix the relative values of the lattice spacing by using eq.~(\ref{necco}), and 
taking as input the central value $a^{-1}(\beta=6)=2.0\gev$. Since our results 
for quark masses are free of leading ${\cal O}(a)$-effects, the first term in 
the extrapolation is of ${\cal O}(a^2)$. For this reason, we extrapolate our 
data to the continuum limit linearly in $a^2$, and obtain the results
\bea
\label{extraV}
&&\ms^{\rm VWI} [\mev] = (102\pm 2) - (2.01\pm 0.25) \, a^2\;, \nonumber \\
&&\nonumber \\
&&\ml^{\rm VWI} [\mev] = (4.20\pm 0.10) - (0.100\pm 0.013) \, a^2\;, \\ 
&& \nonumber \\
&&(\ms/\ml)^{\rm VWI} = (24.25\pm 0.17) + (140 \pm 25) \, a^2\;, \nonumber
\eea
from the VWI, and
\bea
\label{extraA}
&&\ms^{\rm AWI} [\mev] = (106\pm 2) - (1.39\pm 0.28) \, a^2\;,  \nonumber \\
&&\nonumber \\
&&\ml^{\rm AWI} [\mev] = (4.35\pm 0.11) - (0.076\pm 0.014) \, a^2\;, \\
&&\nonumber \\
&&(\ms/\ml)^{\rm AWI} = (24.32\pm 0.17) + (127 \pm 25) \, a^2\;, \nonumber 
\eea
from the AWI, where the lattice spacing must be expressed in units of fm. 
The illustration of the extrapolation of the $\msbar$ strange quark mass to the 
continuum limit is shown in fig.~\ref{fig1}. 
%%%%%%%%%%%%%%%%%%%%%%%%%%%%%%%%%%%%%%%%%%%%%%%%%%%%%%%%%%%%%%%%%%
%%%%%%%%%%%%%%%%%%%%%%%%%%%%%%%%%%%%%%%%%%%%%%%%%%%%%%%%%%%

A pleasant feature of our results is that the quark masses extrapolated to the 
continuum as obtained by using either the AWI or the VWI lead to fully 
consistent determinations. The ${\cal O}(a)$-improved quark masses, with the 
lattice spacings used in our simulations, are very close to the continuum limit.
Although the central values for the quark masses obtained at $\beta=6.45$ are 
slightly smaller than the ones obtained at $\beta = 6.4$, they are completely 
consistent within the statistical errors. 

\section{Systematic uncertainties and final results}  
The errors quoted with the continuum determination of quark masses, in 
eqs.~(\ref{extraV}) and (\ref{extraA}), are statistical only. In this section, 
we discuss the systematic uncertainties and present our final results.
 
The main sources of systematic errors, present in our calculation, are discussed
below.

\vspace{0.3cm}
\noindent
$\circ\ ${\em Additive renormalization of the VWI quark mass}: when using the
VWI method, the renormalized quark mass is obtained from the bare one
by implementing both a multiplicative and an additive renormalization. The
latter is defined by the critical value of the Wilson hopping parameter, 
$\kappa_{cr}$, which, in this study, has been determined from the vanishing of 
the two-point correlation function of the divergence of the axial current. An 
equivalent possibility to fix $\kappa_{cr}$ is to require the pseudoscalar 
meson mass squared to vanish in the chiral limit. The values of light quark 
masses are rather sensitive to the precise choice of $\kappa_{cr}$, and, when 
using this alternative determination, we obtain the results
\bea
\label{extraV2}
&&\ms^{\rm VWI} [\mev] = (110\pm 2) - (1.17\pm 0.28) \, a^2\;, \nonumber \\
&&\nonumber \\
&&\ml^{\rm VWI} [\mev] = (4.57\pm 0.11) - (0.052\pm 0.016) \, a^2\;, \\ 
&& \nonumber \\
&&(\ms/\ml)^{\rm VWI} = (23.94\pm 0.18) + (20 \pm 27) \, a^2\;, \nonumber
\eea
to be compared with those given in eq.~(\ref{extraV}). By combining the
two sets of determinations in eqs.~(\ref{extraV}) and (\ref{extraV2}), we get 
the following estimates of the VWI quark masses
\bea
\label{Vqm}
&&\ms^{\rm VWI} = (106\pm 2\pm 4)\mev \;, \nonumber \\
&&\nonumber \\
&&\ml^{\rm VWI} = (4.38\pm 0.10\pm 0.18)\mev \;, \\ 
&& \nonumber \\
&&(\ms/\ml)^{\rm VWI} = (24.10\pm 0.17\pm 0.15)\;, \nonumber
\eea
where the first error is statistical and the second represents the systematic 
uncertainty due to the spread of the two determinations of $\kappa_{cr}$.

The VWI results in eq.~(\ref{Vqm}) are in perfect agreement with those obtained
in eq.~(\ref{extraA}) by using the AWI method, the differences being smaller 
than 1\%. The systematic uncertainty, however, is larger in the VWI case. For 
this reason, we will quote as our final central values of the results obtained 
from the AWI method, whereas the difference between the two methods will be
included in the systematic error.

%\vspace{0.3cm}
%\noindent
%$\circ\ ${\em Difference between VWI and AWI determinations}: as observed in 
%the previous section, the results for quark masses obtained, in the continuum 
%limit, by using either the VWI or the AWI definition of the bare mass are in 
%good agreement within each other. We then choose to quote, as our final 
%estimate of quark masses, the average between the two determinations, and to 
%include their difference in the systematic error. According to 
%eqs.~(\ref{extraV}) and (\ref{extraA}), this error is of the order of 2\%. It 
%may be due either to residual discretization effects or to a systematic error 
%which affects, in a different way, the calculation of the renormalization 
%constants $Z_S^{-1}$ and $Z_A/Z_P$.

\vspace{0.3cm}
\noindent
$\circ\ ${\em Determination of the lattice spacing}: our estimate of the lattice
spacing has been performed by using eq.~(\ref{necco}), in which the main source 
of uncertainty comes from the input value of the lattice scale at the reference
point $\beta=6$. Our choice $a^{-1}(\beta=6)=2.0(1)\gev$ covers, in a rather 
conservative way, determinations of the scale based on different physical 
quantities, like $f_\pi$, $m_\rho$, $m_{K^*}$, $r_0$, {\it etc.}, which are not 
expected to produce the same estimate of the scale in the quenched 
approximation. In order to evaluate the effect of this uncertainty on the 
determination of the light quark masses, we have repeated the analysis by using 
for $a^{-1}(\beta=6)$ the values 1.9 and 2.1 GeV respectively, and compared the 
results with those given in eqs.~(\ref{extraV}) and (\ref{extraA}) obtained by 
using the central value $a^{-1}(\beta=6)=2.0\gev$. In this way, we find that 
the quark masses vary by approximately 5\% (the mass increases as the 
lattice spacing increases).

\vspace{0.3cm}
\noindent
$\circ\ ${\em Renormalization constants}: a reasonable estimate of the 
systematic uncertainty involved in the non-perturbative $\ri$ calculation of 
the renormalization constants, $Z_A$, $Z_S$ and $Z_P$, can be obtained by 
comparing the results with those obtained by using the chiral Ward identity 
method~\cite{RCS-Roma,LANL,alphaZA}. In the latter case, only scale independent 
quantities can be computed. In particular, the values of the renormalization 
constant $Z_A$ obtained from the two methods are in perfect agreement within 
the statistical errors, while a systematic difference of the order of 5\% is 
observed in the case of the ratio $Z_S/Z_P$. We include this difference in the 
systematic uncertainty.

\vspace{0.3cm}
\noindent
$\circ\ ${\em Finite volume effects}: the spatial extension of the lattices
considered in this study is of the order of 1.6-1.7 fm, which is expected to 
be large enough for finite volume effects to be well under control. In order to
verify this statement, we have performed an independent simulation, at 
$\beta=6.0$, on the volume $24^3 \times 64$, which corresponds (in physical 
units) to a spatial extension of 2.3 fm. The results for the VWI and AWI 
strange quark masses, in the $\msbar$ scheme at the scale $\mu=2\gev$, which are
obtained from the simulation on the larger lattice, are $\ms^{\rm VWI}=85(1)
\mev$ and $\ms^{\rm AWI}=93(3)\mev$ respectively. These results should be 
compared with the values $\ms^{\rm VWI}=82(1)\mev$ and $\ms^{\rm AWI}=92(1)
\mev$ quoted in table~\ref{tab4}. From this comparison, we get an estimate of 
finite volume effects which is of the order of 2\%, which we also account for 
in our final systematic error.

\vspace{0.3cm}
\noindent
$\circ\ ${\em Continuum extrapolation}: the extrapolation of quark 
masses to the continuum limit has been performed by considering only the effect
of a linear term in $a^2$. One may wonder, however, whether higher order 
discretization effects are indeed negligible, particularly for the results 
obtained on the coarsest lattice. For this reason, we have also performed the 
continuum extrapolation without including the point at $\beta=6.0$ in the fit. 
In this way, we find that the results for the strange and the average up/down 
quark masses decrease by 1\% and 2\% respectively.

\vspace{0.3cm}
\noindent
$\circ\ ${\em Perturbative matching}: the conversion factor $R$ of 
eq.~(\ref{conv}), which translates the $\ri$ quark masses to the masses in the 
$\msbar$ scheme, has been computed by using $\alpha_s(m_Z)=0.118$ and the 
number $n_F=4$ of active flavours in the range of scales between 2 and 3 GeV.
By working in the quenched approximation, however, we could have equally
computed this factor by using $n_F=0$ and $\alpha_s$ from $\Lambda_{\rm 
QCD}^{n_F=0}\simeq 0.250$~GeV. By proceeding in this way, we would have 
obtained $R^{(0)}=0.944$, instead of the result $R^{(4)}=0.918$ given in 
eq.~(\ref{conv}). The difference between the two determinations, which is of 
the order of 3\%, represents an intrinsic ambiguity in the quenched 
approximation. We include this difference in the final evaluation of the 
systematic uncertainty.

\vspace{0.3cm}
From the continuum results shown in eq.~(\ref{extraA}), and by adding in
quadrature the systematic uncertainties discussed above, we finally obtain our
best estimates of light quark masses,
\bea
\label{mqfinal}
&&\ms^\msbar (2\gev) = \left( 106\pm 2 \pm 8 \right)\, \mev\;,\cr
&&\cr
&&\ml^\msbar (2\gev) = \left( 4.4\pm 0.1 \pm 0.4 \right) \, \mev\;
\eea
and
\beq
\label{rapfinal}
\ms/\ml = 24.3 \pm 0.2 \pm 0.6 \;,
\eeq
which have been also quoted in the abstract of this paper. Notice that 
most of statistical and systematic uncertainties cancel in the ratio $\ms/\ml$. 

Our results in eqs.~(\ref{mqfinal}) and (\ref{rapfinal}) are in agreement with
the recent extensive lattice QCD calculations of the quark masses performed by
using Wilson fermions in the quenched
approximation~\cite{garden}-\cite{cp-pacs}. They are also in good agreement 
with the current lattice world averages, presented in the 
reviews~\cite{lubicz,Kaneko} and in the 2002 Review of Particle
Physics~\cite{PDG}. The main feature of the present study is, in our opinion, 
the special attention dedicated to the reduction and control of the systematic 
uncertainties within the quenched ap\-pro\-xi\-ma\-tion.

%%%%%%%%%%%%%%%%%%%%%%%%%%%%%%%%%%%%%%%%%%%%%%%%%%%%%%%%%%%%
\section*{Acknowledgments}
%%%%%%%%%%%%%%%%%%%%%%%%%%%%%%%%%%%%%%%%%%%%%%%%%%%%%%%%%%%%
We thank Philippe Boucaud and Guido Martinelli for interesting discussions on 
the subject of this paper. Work partially supported by the European Community's 
Human Potential Programme under HPRN-CT-2000-00145 Hadrons/Lattice QCD.

%%%%%%%%%%%%%  References

\end{document}